# Partition function of 1–, 2–, and 3–D monatomic ideal gas: A simple and comprehensive review

Siti Nurul Khotimah[a,*] dan Sparisoma Viridi[a]

[a]Nuclear Physics and Biophysics Research Division, Faculty of Mathematics and Natural Sciences  
Institut Teknologi Bandung, Jalan Ganeca 10 Bandung 40132, Indonesia

[*]nurul@fi.itb.ac.id

*Abstract*

This article discusses partition function of monatomic ideal gas which is given in Statistical Physisc at Physics Department, Faculty of Mathematics and Natural Sciences, Institut Teknologi Bandung, Indonesia. Students in general are not familiar with partition function. This unfamiliarness was detected at a problem of partition function which was re-given in an examination in other dimensions that had been previously given in the lecture. Based on this observation, the need of a simple but comprehensive article about partition function in one-, two-, and three-dimensions is a must. For simplicity, a monatomic ideal gas is chosen.

*Abstrak*

Artikel ini membahas fungsi partisi Gas ideal monoatomik yang diberikan di kuliah fisika Statistik pada Program Studi Fisika, Fakultas Matematika dan Ilmu Pengetahuan Alam, Institut Teknologi Bandung, Indonesia. Peserta kuliah pada umumnya tidak terbiasa dengan bahasan fungsi partisi. Ketidakbiasaan ini teramati pada soal tentang fungsi partisi ketika diberikan kembali dalam ujian dengan dimensi yang berbeda dari yang telah diberikan sebelumnya dalam kuliah. Berdasarkan pengamatan ini terdapat kebutuhan akan adanya sebuah artikel mengenai fungsi partisi dalam kasus satu-, dua-, dan tiga-dimensi. Gas ideal monoatomik dipilih agar sederhana.

**Kata Kunci**: *fisika statistik, fungsi partisi, gas ideal monoatomik*

## 1. Introduction

Partition function for monatomic ideal gas is commonly discussed for three-dimensional case [1], but it is also interesting, in analogy and mathematical point of view, to discuss it in one- or two-dimension. Partition function can be viewed as volume in n-space occupied by a canonical ensemble [2], where in our case the canonical ensemble is the monatomic ideal gas system.

In order to understand this work reader must already familiar with $\Gamma$-integral and its relation with factorial *n*! [3].

## 2. Theory

In general, a system of particles which obeys Maxwell-Boltzmann statistics, has a definition for partition function as:

$$Z = \sum_j g_j\, e^{-\varepsilon_j / kT} \,. \quad (1)$$

When there are $\Delta G_j$ energy states within the macrolevel then Equation (1) will turn into

$$Z = \sum_j \Delta G_j\, e^{-\varepsilon_j / kT} \,. \quad (2)$$

Energy of each particle, using the principle of quantum mechanics for single particle in a box, is given by [4]

$$\varepsilon_j = \frac{h^2 V^{-2/3}}{8m} n_j^{\,2} \,. \quad (3)$$

For 3-D case as illustrated in Figure 1(a), it can be written that

$$G_j^{(3)} = \frac{1}{8} \cdot \frac{4}{3} \pi n_j^3, \qquad (4)$$

then number of states of particles which have quantum number between $n_j$ and $n_j + \Delta n_j$ or have energy between $\varepsilon_j$ and $\varepsilon_j + \Delta \varepsilon_j$ is

$$\Delta G_j^{(3)} = \frac{1}{2} \pi n_j^2 \Delta n_j. \qquad (5)$$

Substitution Equation (5) into Equation (2) will give:

$$Z^{(3)} = \frac{1}{2} \pi \sum_j n_j^2 \Delta n_j\, e^{-\varepsilon_j/kT}. \qquad (6)$$

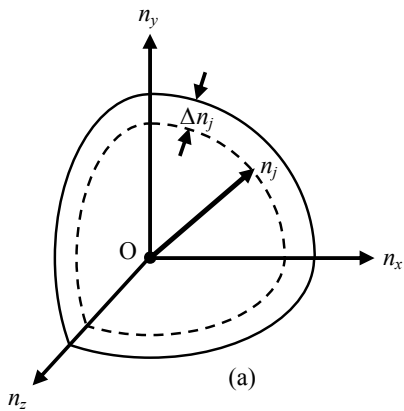

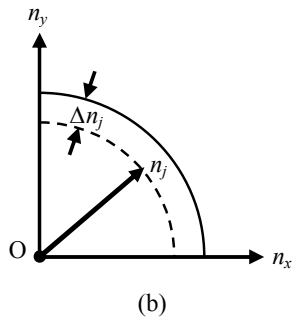

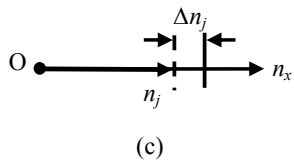

Figure 1. n-space for case of: (a) one-, (b) two-, and (c) three-dimensional monatomic ideal gas.

By using Equation (3) into Equation (6) and write the result in continuous form

$$Z^{(3)} = \frac{\pi}{2} \int_0^\infty n_j^2 \exp\!\left(-\frac{h^2 V^{-2/3}}{8mkT} n_j^2\right) dn_j. \qquad (7)$$

The $\Gamma$-integral and its relation will give immediately the result of Equation (7) in a form of

$$Z^{(3)} = V\left(\frac{2\pi mkT}{h^2}\right)^{3/2}. \qquad (8)$$

Then, the next is 2-D case as illustrated in Figure 1(b). This case will give:

$$G_j^{(2)} = \frac{1}{4} \cdot \pi n_j^2, \qquad (9)$$

then number of state of particles which have quantum number between $n_j$ and $n_j + \Delta n_j$ or have energy between $\varepsilon_j$ and $\varepsilon_j + \Delta \varepsilon_j$ is

$$\Delta G_j^{(2)} = \frac{1}{2} \pi n_j \Delta n_j. \qquad (10)$$

Substitution Equation (10) into Equation (2) will give

$$Z^{(2)} = \frac{1}{2} \pi \sum_j n_j \Delta n_j\, e^{-\varepsilon_j/kT}. \qquad (11)$$

Following previous steps for 3-D case, but by using $A^{-1}$ instead of $V^{-2/3}$, we can arrive at:

$$Z^{(2)} = \frac{\pi}{2} \int_0^\infty n_j \exp\!\left(-\frac{h^2 A^{-1}}{8mkT} n_j^2\right) dn_j, \qquad (12)$$

This gives

$$Z^{(2)} = A\left(\frac{2\pi mkT}{h^2}\right)^{2/2}. \qquad (13)$$

Finally by following the similar steps we can obtain that for 1-D case as illustrated in Figure 1(c)

$$G_j^{(1)} = n_j, \qquad (14)$$

then number of state of particles which have quantum number between $n_j$ and $n_j + \Delta n_j$ or have energy between $\varepsilon_j$ and $\varepsilon_j + \Delta \varepsilon_j$ is

$$\Delta G_j^{(1)} = \Delta n_j. \qquad (15)$$

Substitution Equation (15) into Equation (2) will give

$$Z^{(1)} = \sum_j \Delta n_j \, e^{-\varepsilon_j/kT}. \quad (16)$$

Then following similar steps for 3- and 2-D case but by using $L^{-2}$ instead of $V^{-2/3}$, it is obtained that:

$$Z^{(1)} = \int_0^\infty \exp\left(-\frac{h^2 L^{-2}}{8mkT} n_j^2\right) dn_j, \quad (17)$$

This gives

$$Z^{(1)} = L\left(\frac{2\pi mkT}{h^2}\right)^{1/2}. \quad (18)$$

Equation (7), (12), and (17) can be solved using the following relations [3]

$$\Gamma(n+1) = \int_0^\infty x^n e^{-x} dx = n\Gamma(n) = n!, \quad (19)$$

$$\Gamma(1) = 1, \quad (20)$$

$$\Gamma\left(\frac{1}{2}\right) = \sqrt{\pi}, \quad (21)$$

and

$$\int_0^\infty x^n e^{-ax^2} dx = \frac{1}{2a^{(n+1)/2}} \Gamma\left(\frac{n+1}{2}\right). \quad (22)$$

## 3. Results and Discussions

It can be seen from Equation (8), (13), and (18) that there is a regularity in writing the partition function of monatomic ideal gas for 1-, 2-, and 3-D case as shown in Table 1.

Table 1. Comparison of partition function of monatomic ideal gas for 1-, 2-, and 3-D case.

| Case | Partition function |
|------|---------------------|
| 1-D | $Z^{(1)} = L\left(\dfrac{2\pi mkT}{h^2}\right)^{1/2}$ |
| 2-D | $Z^{(2)} = A\left(\dfrac{2\pi mkT}{h^2}\right)^{2/2}$ |
| 3-D | $Z^{(3)} = V\left(\dfrac{2\pi mkT}{h^2}\right)^{3/2}$ |

We can then say that the partition function of monatomic ideal gas can be written in general form, which is

$$Z^{(D)} = L^D \left(\frac{2\pi mkT}{h^2}\right)^{D/2}, \quad (23)$$

where $D$, which is the dimension, can have value of 1, 2, or 3. Then thermodynamics property that similar to pressure $p$ in 3-D case can also be defined in 2- and 1-D case. This property usually derived from Helmholtz free energy $F$, which is related to partition function through

$$F^{(D)} = -NkT \ln Z^{(D)} \quad (24)$$

and then

$$p^{(D)} = -\left[\frac{\partial F^{(D)}}{\partial L^D}\right]_T = NkT\left[\frac{\partial \ln Z^{(D)}}{\partial L^D}\right]_T. \quad (25)$$

Table 2. Comparasion of pressure-like thermodynamics properties for monatomic ideal gas for 1-, 2-, and 3-D case.

| Case | Pressure-like properties | Unit (SI) |
|------|--------------------------|-----------|
| 1-D | $p^{(1)} \equiv F = \dfrac{NkT}{L}$ | kg·m·s$^{-2}$ |
| 2-D | $p^{(2)} \equiv \tau = \dfrac{NkT}{A}$ | kg·s$^{-2}$ |
| 3-D | $p^{(3)} \equiv p = \dfrac{NkT}{V}$ | kg·m$^{-1}$·s$^{-2}$ |

We can then write a bit more generality the equation in Table 2 as follows:

$$p^{(D)} = \frac{NkT}{L^D} = \frac{NkT}{V^{(D)}} \quad (26)$$

Where $L \equiv L^{(1)}$ and $V^{(D)} = L^D$ is the volume of the D-dimensional domain occupied by the gas. Physically, $p^{(1)} \equiv F$ is a just a "force" exerted at the two endpoints of a segment of length $L$; $p^{(2)} \equiv \tau$ is the surface tension exerted along the closed line which is the boundary of the area $A$ where the gas is confined; and $p^{(3)} \equiv p$ is the well-known outward-pointing pressure of the gas, exerted at its two dimensional boundary.

Table 3. Equation of state for monatomic ideal gas for 1-, 2-, and 3-D case.

| Case | Equation of state |
|------|-------------------|
| 1-D | $FL = NkT$ |
| 2-D | $\tau A = NkT$ |
| 3-D | $pV = NkT$ |

Table 3 shows us the equation of state of monatomic ideal gas for 1-, 2-, and 3-D case. The 3-D case is the most familiar form for the students, while this form is already taught since at senior high school.

Problems usually arise when the students do not understand where Equation (4), (9), and (14) are originated from and also why Equation (3) is needed. In this case the role of lecture is very important to guide them in the lecture.

It can be illustrated for 3-D case that Equation (4) is actually $\frac{1}{8}$ volume of a sphere with radius $n_j$ and Equation (5) is $\frac{1}{8}$ volume of a shell of the sphere with radius $n_j$ and thickness $\Delta n_j$. Figure 1(a) shows the illustration. The number $\frac{1}{8}$ appears since we consider only positive value of $n_x$, $n_y$, and $n_z$, which lies only in one octane or $\frac{1}{8}$ of total volume of the sphere, as this approach of explanation is suggested [1].

Following the approach for 2-D case as illustrated in Figure 1(b), Equation (9) is area with positive value of $n_x$ and $n_y$ and radius $n_j$, and Equation (10) is area of a ring with radius $n_j$ and thickness $\Delta n_j$. The number $\frac{1}{4}$ arise since we consider only one quadrant or $\frac{1}{4}$ or the total area.

Then, finally for 1-D case as illustrated in Figure 1(c), Equation (14) is length of $n_j$ and Equation (15) is a region with thickness $\Delta n_j$. Since $n_x$ is replaced by $n_j$ then it has already only positive value.

So, perhaps it is also necessary to give an common picture about relation of $n_j$ with $n_x$, $n_y$, and $n_z$ in case of 1-, 2-, and 3-D case and also the factor 1, $\frac{1}{4}$, and $\frac{1}{8}$. Table 4 gives the illustration.

Table 4. Expression of $n_j$ and the factor in front of Equation (5), (9), and (14) for 1-, 2-, and 3-D case respectively.

| Case | $n_j$ | Factor |
|---|---|---|
| 1-D | $n_j^2 = n_x^2$ | 1 |
| 2-D | $n_j^2 = n_x^2 + n_y^2$ | $\frac{1}{4}$ |
| 3-D | $n_j^2 = n_x^2 + n_y^2 + n_z^2$ | $\frac{1}{8}$ |

## 4. Conclusion

The partition function of monatomic ideal gas system for 1-, 2-, and 3-D case has been reviewed and compared. General formulation has also been shown. Brief and simple explanation how theses partition functions derived is also given in order to help the students to understand it with olny few efforts.

## Acknowledgments

Authors would like to thank to Raul A. Simon (Santiago, Chile) for his valuable comment on the the physical interpretation of "pressure like properties" of the different dimensional cases. Authors also would like to thank to Kelompok Pendidikan Fisika at Physics Department, Insitut Teknologi Bandung for the friendly discussion atmosphere and Hibah Kapasitas FMIPA.PN-6-18-2010 for partially support to this work.